\documentclass[11pt,reqno]{amsart}
\usepackage[T1]{fontenc}
\usepackage{graphicx, amssymb, amsmath}
\usepackage{cite}
\usepackage{color}
\definecolor{MyLinkColor}{rgb}{0,0,0.4}

\newcommand{\R}{{\mathbb R}}

\newcommand{\N}{{\mathbb N}}

\newcommand{\cO}{\mathcal{O}}

\newcommand{\cH}{\mathcal{H}}

\newcommand{\ov}{\overline}
\newcommand{\p}{\partial}

\newcommand{\0}{\Omega}
\newcommand{\G}{\Gamma}

\newtheorem{thm}{Theorem}[section]

\theoremstyle{remark} 
\newtheorem{rem}[thm]{Remark}

\numberwithin{equation}{section} 
\usepackage[colorlinks=true,linkcolor=MyLinkColor,citecolor=MyLinkColor]{hyperref}

\begin{document}

\title[Recovery of steady water waves from the horizontal velocity on a line of symmetry]{Recovery of traveling water waves with smooth vorticity from the horizontal velocity on a 
line of symmetry for various wave regimes}
\thanks{Partially supported by DFG Research Training Group~2339 ``Interfaces, Complex Structures, and Singular Limits in Continuum Mechanics - Analysis and Numerics''}
\author{Daniel B\"ohme}
\author{Bogdan-Vasile Matioc}
\address{Fakult\"at f\"ur Mathematik, Universit\"at Regensburg \\ D--93040 Regensburg, Deutschland}
\email{daniel.boehme@ur.de}
\email{bogdan.matioc@ur.de}

\begin{abstract}
In the general context of rotational water waves  with a smooth vorticity
 it is shown that the wave profile can be recovered from the  horizontal component of the  velocity field on a line of symmetry.
 The method, which applies to waves of finite and infinite depth, uses only  the values of the horizontal velocity of particles located on the line of symmetry that are close to the wave surface.
 In fact, together with the wave surface we recover also  the velocity field  in a suitable surface layer. 
The explicit recovery formula is valid under the assumption that there are no stagnation points in the fluid  for both periodic and solitary waves in each of the three regimes of  
gravity, capillary-gravity, and capillary waves.
The efficiency of this method is illustrated in the context of the explicit   solutions provided by Crapper for periodic capillary waves and Gerstner for periodic gravity waves. 
\end{abstract}

%%% NEED TO ADD MSC %%%

\maketitle

\pagestyle{myheadings}
\markboth{\sc{D. B\"ohme \& B.-V.~Matioc}}{\sc{Recovery of traveling water waves from the horizontal velocity on a line of symmetry}}

%%%%%%%%%%%%%%%%%%%%%%%%%%%%%%%%%%%%%%%%%%%%%%%%%%
%%%%%%%%%%%%%%%%%%%%%%%%%%%%%%%%%%%%%%%%%%%%%%%%%%
%%%%%%%%%%%%%%%%%%%%%%%%%%%%%%%%%%%%%%%%%%%%%%%%%%
%%%%%%%%%%%%%%%%%%%%%%%%%%%%%%%%%%%%%%%%%%%%%%%%%%
 \section{Introduction}\label{Sec:1}
%%%%%%%%%%%%%%%%%%%%%%%%%%%%%%%%%%%%%%%%%%%%%%%%%%
%%%%%%%%%%%%%%%%%%%%%%%%%%%%%%%%%%%%%%%%%%%%%%%%%%
%%%%%%%%%%%%%%%%%%%%%%%%%%%%%%%%%%%%%%%%%%%%%%%%%%
%%%%%%%%%%%%%%%%%%%%%%%%%%%%%%%%%%%%%%%%%%%%%%%%%%

In this paper we show that the wave profile of two-dimensional rotational symmetric waves with smooth vorticity can be recovered from the horizontal velocity of the fluid particles 
on a line of symmetry of the flow.
More precisely, we prove that the horizontal velocity of the particles located  on the symmetry line in a  (arbitrary thin) surface layer identifies
 the wave profile and the fluid velocity in each point of the surface layer.
This property is valid regardless of whether the waves propagate in shallow or deep water, whether they are periodic or solitary, and also it 
does not distinguish between gravity waves, capillary waves, or capillary-gravity waves. 
It is important to mention that though at a formal level stagnation points need to be excluded  only in a thin surface layer below the wave profile and on the line of symmetry, 
 the justification of the recovery method is build on the assumption that the entire flow does not contain  stagnation points, see Remark~\ref{R:1}.
 This aspect is exemplified in Figure~\ref{Figure0}  in the context of Gerstner's cycloidal wave, which has  stagnation points at the crests, 
 where the wave profile  is recovered by considering the trough line as a line of symmetry.

From a mathematical point of view, we deal with an  elliptic boundary value problem (see e.g. the stream function formulation~\eqref{SF}) defined in a fluid domain which is unknown
since, on the one hand,  the wave surface is unknown 
and, on the other hand, because we do not distinguish between waves of finite depth and waves of infinite depth (boundary conditions on the bed or/and 
at infinity play no role in the analysis and are therefore disregarded).
Our main result stated in Theorem~\ref{MT} shows that, knowing some information about the solution to the problem,
 we may reconstruct the wave surface regardless of the values of the Earth's gravity  constant $g\geq0$ and of the surface tension coefficient $\sigma\geq0$.  
  Our result relies to a large extend on the fact that the height function $h$
  associated to waves without stagnation points, see Section~\ref{Sec:22}, is smooth with respect to the horizontal variable~$q$ and satisfies  the a priori estimates~\eqref{ana23}, see \cite{CLW13, BM13}.  
The structure of the  quasilinear elliptic equation satisfied by~$h$, see~\eqref{eq:hod}$_1$, is a further building block of our recovery method.
  This paper extends the previous result established in \cite{BM15} in the particular context of periodic gravity waves of finite depth also to other wave regimes and exemplifies the recovery method 
  in the context  of Crapper's exact solutions  for periodic capillary waves and Gerstner's periodic gravity waves (with and without stagnation points), see Section~\ref{Sec:3}.
  For an extension of the method from \cite{BM15} to the setting of stratified waves we refer to \cite{XYL21}.

The question of determining the free surface of steady water waves from measured flow data is 
of great importance in the field of fluid dynamics and at the same time a challenging mathematical problem.
Since direct measurements of  the  surface of ocean waves are difficult and costly, alternative ways of computing the free surface from  measured flow data are of great importance.
Theorem~\ref{MT} is a contribution in this direction since the velocity on crest or trough lines can be  measured in numerical but also laboratory experiments, see \cite{GCHJ03, HHY12, CKS15,  SCJ01}.
In this context, a further important direction of research addresses the question whether the wave profile can be recovered from the measured pressured data on the fluid bed.
This research considers mainly the setting of gravity waves of finite depth which are either irrotational or have a constant vorticity.
A positive answer to this question was provided in \cite{DH14An} in the general setting of solitary water waves with  analytic vorticity, where it was proven that the
pressure function on the flat bed identifies the surface profile in a unique way.  
The result of \cite{DH14An} was improved in \cite{CW18}  where less regularity was required for the  vorticity of the flow. 
 Explicit formulas that relate the pressure on the bed to the wave profile were derived first mainly in the
linear setting, cf. \cite{ES08, Tsai05, BL95, KC94}, and  involve   so called pressure transfer functions.
However, this approach   does  not take into account nonlinear effects, which leads to   prediction errors.
Explicit and exact nonlinear and nonlocal formulas, obtained from the Euler formulation of the water-wave problem without approximation, that 
relate the wave profile  to the pressure on the flat bed were derived  in the irrotational setting in  \cite{De11, OD12, C12, Hsu14}.
Further exact formulas, more tractable from a numerical point of view, were provided for steady periodic waves in \cite{Cl13, CC13}.
As recently shown in \cite{CD20}, the  formulas from \cite{Cl13, CC13} are applicable  also to the celebrated extreme Stokes wave which possesses a stagnation point at the crest.
A recovery formula in the more general setting of rotational waves with constant vorticity, and possibly with stagnation points, was derived only recently in \cite{CLH23} 
(see also \cite{HT18} for recovery  formulas for linear and weakly nonlinear rotational water waves with  arbitrary vorticity distributions).

%%%%%%%%%%%%%%%%%%%%%%%%%%%%%%%%%%%%%%%%%%%%%%%%%%
%%%%%%%%%%%%%%%%%%%%%%%%%%%%%%%%%%%%%%%%%%%%%%%%%%
%%%%%%%%%%%%%%%%%%%%%%%%%%%%%%%%%%%%%%%%%%%%%%%%%%
%%%%%%%%%%%%%%%%%%%%%%%%%%%%%%%%%%%%%%%%%%%%%%%%%%
 \section{The mathematical problem and the statement of the main result}\label{Sec:2}
%%%%%%%%%%%%%%%%%%%%%%%%%%%%%%%%%%%%%%%%%%%%%%%%%%
%%%%%%%%%%%%%%%%%%%%%%%%%%%%%%%%%%%%%%%%%%%%%%%%%%
%%%%%%%%%%%%%%%%%%%%%%%%%%%%%%%%%%%%%%%%%%%%%%%%%%
%%%%%%%%%%%%%%%%%%%%%%%%%%%%%%%%%%%%%%%%%%%%%%%%%%

We consider the setting of smooth two-dimensional steady waves traveling at the surface of an inviscid  and incompressible fluid of constant density, which we regard as being water,   
with  constant  wavespeed $c>0$.
We choose the Cartesian coordinates $(x,y,z)$ such that the~$x$-axis is the direction of wave propagation and  the $y$-axis points vertically upwards, 
 the  flow being independent of the  $z$-variable.
Under these considerations, the equations governing the motion in the bulk are the Euler equations.
Since we consider the framework  of steady waves, in a reference frame moving with the wavespeed the Euler equations are encompassed by the system
\begin{equation}\label{Eq1}
\left.
\arraycolsep=1.4pt
\begin{array}{rclc}
u u_x+vu_y&=&-P_x,\\[1ex]
uv_x+vv_y&=&-P_y-g,\\[1ex]
u_x+v_y&=&0,
\end{array}
\right\}
\end{equation}
where $(u,v)$ is the velocity field and $P$ is the pressure observed in the moving reference frame, while $g\geq0$  is the Earth's gravity constant (which is set to be zero when gravity effects are neglected).
The density of the fluid is assumed to be constant $1$.
These equations are  satisfied in the fluid domain $\Omega_\eta,$ which is bounded from above by the the wave surface  
\[\Gamma_\eta:=\{(x,\eta(x))\,:\, x\in\R\}.\]
We consider both waves of finite and infinite depth.
In the first case $\0_\eta$ is  bounded from below by a flat impermeable bed, in the second case  $\0_\eta$  is unbounded from below.
Letting~${\sigma\in[0,\infty)}$  denote   the surface tension coefficient at the water surface (surface tension effects are neglected when $\sigma=0$ ), we supplement \eqref{Eq1} 
by the following natural   dynamic and the kinematic boundary conditions at the wave surface
\begin{equation}\label{Eq2}
\left.
\arraycolsep=1.4pt
\begin{array}{rclc}
P&=&-\sigma\cfrac{\eta''}{(1+\eta'^2)^{3/2}},\\[2ex]
v&=&u\eta'
\end{array}
\right\}\quad\text{on $\G_\eta$}.
\end{equation}
Depending on the physical scenario which is considered, these equations have to be supplemented by periodicity assumptions on the flow and/or   boundary conditions which 
describe the asymptotic behavior of  the flow at infinity and/or on the bed.
 As these do not play any role in our analysis, we refrain from  mentioning them here.

We consider solutions to \eqref{Eq1}-\eqref{Eq2} for which the flow does not contain any stagnation points, that is
 \begin{equation}\label{Eq3}
\arraycolsep=1.4pt
\begin{array}{rclc}
\sup u&<&0\quad\text{in $\0_{\eta}$}.
\end{array}
\end{equation}
We emphasize that   the vorticity 
\[
\omega:=u_y-v_x
\]
of the flow is not subject to any restrictions (apart from the regularity assumptions \eqref{RA} below).
Finally, we shall consider only waves which are symmetric with respect to a vertical line (which may be a crest or trough line).
Since the equations of motion are translation invariant with respect to the $x$-variable, we may take this line as being the vertical line~${\{x=0\},}$ hence we assume that 
 \begin{equation}\label{sym}
\arraycolsep=1.4pt
\left.
\begin{array}{rclc}
\eta(x)&=&\eta(-x)\quad\text{for all $x\in\R$},\\[1ex]
(u,v,P)(x,y)&=&(u,-v,P)(-x,y)\quad\text{for all $(x,y)\in\0_\eta.$}
\end{array}
\right\}
\end{equation}
We point out that there exists vast body of literature which investigates the symmetry of rotational and irrotational waves, see e.g.  \cite{CoEhWa07, Co-Es04_1,CoEs04_b, CS88, hur-08_1, hur-07_1, Eh07, MatB14, TO00, HMa14, WalS09}.

We  restrict our attention to the setting of classical solutions to  problem  \eqref{Eq1}-\eqref{Eq3} which posses a smooth vorticity, that is we assume that 
 \begin{equation}\label{RA}
u,\,v,\, P\in {\rm BUC}^{1+\alpha}( \Omega_{\eta}),\quad \eta\in{\rm BUC}^{2+\alpha}(\mathbb{R}),\quad \omega\in {\rm BUC}^{\infty}(\0_\eta),
 \end{equation}
where $\alpha\in(0,1) $ is fixed.
Given $k\in\N$ and an open set $\cO\subset\R^n$, $n\geq 1$, the Banach space~${{\rm BUC}^{k+\alpha}(\cO)}$  consists of the functions which have bounded derivatives up to order~$k$ and
 uniformly $\alpha$-H\"older continuous derivatives of order $k$.
Moreover, we define 
$${\rm BUC}^{\infty}(\cO) =\bigcap_{k\in\N} {\rm BUC}^{k+\alpha}(\cO).$$
Before stating our result, we  first reexpress the problem~\eqref{Eq1}-\eqref{Eq3}, under the assumptions~\eqref{sym} and~\eqref{RA}, 
by using the so-called stream function  and the height function. 
 
 \subsection{The stream function formulation}\label{Sec:21}
The stream function $\psi:\ov{\Omega_{\eta}}\to\R$ is defined by the formula
\[
\psi(x,y):=-\int_y^{\eta(x)}u(x,s)\, {\rm d}s \qquad \text{for $(x,y)\in\ov{\0_\eta}$.}
\]
 The function $\psi$ vanishes at the wave surface and $\psi(x,y)=\psi(-x,y)$ for all~${(x,y)\in\0_\eta}$.
Combining ~\eqref{Eq1}$_3$, \eqref{Eq2}$_2$, and~\eqref{RA}, we deduce that~$\nabla \psi=(-v,u)\in  {\rm BUC}^{1+\alpha}( \Omega_{\eta})^2$.
In particular, since $\psi_y=u<0$, we deduce that $\psi$ takes only positive values below the surface.
  Recalling~\eqref{Eq3} and~\eqref{RA}, the implicit function theorem ensures  that for each constant~${\mu\geq 0}$, the level set~${\{\psi(x,y)=\mu\}}$ 
  -- which is a streamline of the steady flow --
  is the graph of an even function that belongs to ${\rm BUC}^{2+\alpha}(\mathbb{R})$.

We next introduce the Dubreil-Jacotin's semihodograph transform~${\mathcal{H}:=(q,p):\ov{\0_{\eta}}\to\ov{\0}}$, with~$\0:=\cH(\0_\eta),$ 
by the formula
\[
\cH(x,y):=(x,-\psi(x,y)).
\]
In the setting of waves of infinite depth   it holds that $\0:=\{(q,p)\in\R^2\,:\, p<0\}$, respectively, for waves of finite depth,   we have  $\0:=\{(q,p)\in\R^2\,:\,p_0< p<0\}$,
where $-p_0$ is the value taken by $\psi$ on the fluid bed.
In view of \eqref{Eq3}, we deduce that $\cH$ is a ${\rm C}^{2+\alpha}$-diffeomorphism from $\0_{\eta}$ to $\0$, which maps the streamlines of the steady flow onto horizontal lines in $\0$.
Observing that
\begin{equation*}
\frac{\partial(q,p)}{\partial(x,y)}=
\begin{pmatrix}
 1&0\\
 v&-u
\end{pmatrix}
\qquad\text{and}\qquad \frac{\partial (x,y)}{\partial(q,p)}\circ\mathcal{H}=\begin{pmatrix}
 1&0\\
 \cfrac{v}{u}&-\cfrac{1}{u}
\end{pmatrix},
\end{equation*}
we compute 
\begin{align*}
 \frac{d}{dq} \big(\omega\circ \mathcal{H}^{-1}\big)=\frac{u(u_{xy}-v_{xx})+v(u_{yy}-v_{xy})}{u}\circ\mathcal{H}^{-1} \qquad\text{in $\0$}.
\end{align*}
Moreover, differentiating  \eqref{Eq1}$_1$ with respect to $y$ and \eqref{Eq1}$_2$ with respect to $x$, we obtain, after taking the difference 
 of the resulting identities that
\begin{align*}
 0=&\big(uu_x+vu_y\big)_y-\big(uv_x+vv_y\big)_x\\[1ex]
 =&\big[u(u_{xy}-v_{xx})+v(u_{yy}-v_{xy})\big]+\omega(u_x+v_y)\quad\text{in $\0_\eta$}.
\end{align*}
Using also the incompressibility condition \eqref{Eq1}$_3$, we conclude that 
\begin{align*}
 \frac{d}{dq} \big(\omega\circ \mathcal{H}^{-1}\big)=0\quad\text{in $\0$.}
\end{align*}
Therefore, the vorticity function $\gamma:=\omega\circ \mathcal{H}^{-1}$ depends only on the variable $p$, that is~$\gamma=\gamma(p)$ and 
\[
\omega(x,y)=\gamma\circ \mathcal{H}(x,y)=\gamma(-\psi(x,y))\qquad\text{for $(x,y)\in\ov{\0_\eta}$.}
\]
Finally, defining the total energy  of the flow $E:\ov{\0_\eta}\to\R$ by the formula 
\begin{equation}\label{Bern}
E(x,y):= \frac{u^2+v^2}{2}(x,y)+gy+P(x,y)-\int_0^{\psi(x,y)}\gamma(-s)\,{\rm  d}s,
\end{equation}
partial differentiation together with \eqref{Eq1}$_1$-\eqref{Eq1}$_2$ shows that $\nabla E=0$ in $\0_\eta$.
Herewith we recover Bernoulli's law which states that $E$ is constant in $\0_\eta$. 
 Evaluating \eqref{Bern} at the wave surface, we obtain from \eqref{Eq2}$_1$ and 
 \[
\Delta\psi=\p_x\psi_x+\p_y\psi_y=-v_x+ u_y=\omega=\gamma(-\psi) \quad\text{in $\0_\eta$}, 
 \]
 that $(\psi,\eta)$ solves the system
\begin{equation}\label{SF}
\left.
\arraycolsep=1.4pt
\begin{array}{rclc}
\Delta\psi&=&\gamma(-\psi)\quad\text{in $\0_\eta$},\\[1ex]
\psi&=&0\quad\text{on $\G_\eta$},\\[1ex]
|\nabla\psi|^2+2g\eta-2\sigma\cfrac{\eta''}{(1+\eta'^2)^{3/2}}&=&Q\quad\text{on $\G_\eta$},
\end{array}
\right\}
\end{equation}
where $Q$ is a constant. 
The conditions on the periodicity of the flow and/or the  boundary conditions prescribing  the behavior of the flow at infinity and/or on the bed are again not explicitly mentioned.

\subsection{The height function formulation}\label{Sec:22}
The height function $h:\ov \Omega\to\R$ is defined by the formula
\[
h(q,p):=y\circ \cH^{-1}(q,p) 
\]
and specifies  the vertical position of the fluid particle $(x,y)=\cH^{-1}(q,p)$.
The height function~$h$ satisfies $\nabla h\in {\rm BUC}^{1+\alpha}(\0)$, is symmetric with respect to the vertical line $\{q=0\}$, that is $h(q,p)=h(-q,p)$ for all $(q,p)\in\0,$ and 
 solves the following  system
\begin{equation}\label{eq:hod}
\left.
\begin{array}{rllll}
(1+h_q^2)h_{pp}-2h_ph_qh_{pq}+h_p^2h_{qq}-\gamma h_p^3&=&0&\text{\qquad in $\Omega$},\\[1ex]
1+h_q^2+(2gh-Q)h_p^2-\displaystyle 2\sigma\frac{h_p^2 h_{qq}}{(1+h_q^2)^{3/2}}&=&0&\text{\qquad on $p=0$},\\
\end{array}
\right\}
\end{equation}
supplemented by  periodicity conditions and/or  appropriate boundary conditions  on $h$.
It is important to mention that, given any streamline of the flow, there exists $p\leq 0$ such that~${h(\cdot,p)}$ is a parameterization of this streamline (the wave surface corresponds to the choice $p=0$).

\subsection{The main result}\label{Sec:23}
The main result of this paper is formulated in Theorem~\ref{MT} below and states that for  periodic or solitary waves of finite or infinite depth,  which posses a vertical line of symmetry, the wave profile 
and the flow in any given surface layer are determined by the horizontal velocity on the segment of the line of symmetry which is contained in the given layer. 
This holds regardless of the values of $g\geq 0$ and $\sigma\geq0. $

\begin{thm}\label{MT}
Let $(u,v,P,\eta)$ be a solution to \eqref{Eq1}-\eqref{Eq3} satisfying \eqref{sym}  and \eqref{RA}, with constants $g,\, \sigma\geq 0$ such that 
\[
g+\sigma>0.
\]
Assume further that the values $\mathfrak{u}(y):=u(0,y)$ of $u$ on the line of symmetry $\{x=0\}$ are known for all $y\in[\overline{y},\eta(0)],$ where the point $(0,\overline y)$ lies in the fluid domain $\overline{\0_\eta}$ below the wave surface.
We  further  set~${\ov p:=-\psi(0,\ov y)}$ and we assume  there exists a constant $L>0$ such that the corresponding  height function $h$ satisfies
\begin{equation}\label{ana23}
\|\p_q^mh\|_{{\rm BUC}^{2+\alpha}(\R\times (\ov p,0))}\leq L^m m!\qquad\text{for all $m\geq 0$.} 
\end{equation} 

Then the free wave surface and the velocity field in the surface layer $$\{(x,y)\in\ov{\0_\eta}\,:\, 0\leq \psi(x,y)\leq -\ov p\}$$ are determined by the function $[y\mapsto \mathfrak{u}(y)]:[\overline{y},\eta(0)]\to\R.$
\end{thm}

\begin{rem}\label{R:1}
Though we formulate  \eqref{ana23} as a condition on the solution to \eqref{eq:hod}, classical solutions to the water wave problem which do not contain stagnation points a priori satisfy this estimate,  see \cite{CLW13, BM13}.
\end{rem}

We are now in a position to prove Theorem~\ref{MT}.
\begin{proof}
As a direct consequence of \eqref{ana23}, for every $p\in[\ov p,0]$, the map $[q\mapsto h(q,p)]:\R\to\R$ is real-analytic.
Indeed, given $p\in[\ov p,0]$ and $q_0\in\R$, Taylor's theorem together with \eqref{ana23} yields
\begin{equation}\label{Taylor}
\Big|h(q,p)-\sum_{k=0}^n\frac{\p_q^kh(q_0,p)}{k!}(q-q_0)^k\Big|\leq \frac{\|\p_q^{n+1}h(\cdot,p)\|_\infty}{(n+1)!}|q-q_0|^{n+1}\leq (L|q-q_0|)^{n+1},
\end{equation}
which shows that the Taylor series converges on  $(q_0-L^{-1},q_0+L^{-1}) $.
In particular, the wave surface $\eta=h(\cdot,0)$ and also the streamline $h(\cdot , \ov p)$, are real-analytic.
Since  the stream function~$\psi$  solves the elliptic Dirichlet boundary value problem
\begin{equation*}
\left.
\arraycolsep=1.4pt
\begin{array}{rclc}
\Delta\psi&=&\omega\quad\text{in $\{h(x, \ov p)< y<\eta(x)\}$},\\[1ex]
\psi&=&0\quad\text{on $\G_\eta$},\\[1ex]
\psi&=&\ov p\quad\text{on $\{y=h(x,\ov p)\}$} 
\end{array}
\right\}
\end{equation*}
in a smooth domain and with $\omega$ smooth, standard elliptic theory, see e.g. \cite[Theorem~6.19]{GT01}, ensures that $\psi $ is smooth (up to the boundary) in $\{h(x, \ov p)< y<\eta(x)\}$. 
Recalling the definition of the vorticity function $\gamma$, we deduce that $\gamma\in{\rm C}^\infty([\ov p,0]).$ Moreover, also $h$ is smooth  (up to the boundary) in  the strip 
$\{\ov p<p<0\}$.

Using now the symmetry  of $h$ with respect to the vertical line $\{q=0\},$ we deduce that~${\p_q^nh(0,p)=0}$ for all $p\in[\ov p,0]$ and all odd integers $n\in\N,$ hence
\[
h(q,p)=\sum_{n=0}^\infty a_{2n}(p)q^{2n}\qquad \text{for all $|q|<L^{-1}$ and $p\in[\ov p,0]$,}
\]
where $a_{2n}\in {\rm C}^\infty([\ov p,0]),$ $n\in\N$, and 
\[
a_0(p)=h(0,p),\qquad p\in[\ov p,0].
\]
We next prove that the equation \eqref{eq:hod}$_1$ enables us to express all coefficient functions $a_{2n}$ with~$n\geq 1$ in terms of  (the derivatives of) $a_0$ and $\gamma$. 
Indeed, arguing as in the proof  of~\eqref{Taylor}, we infer from~\eqref{RA} that for all $|q|<L^{-1}$ and $p\in[\ov p,0]$
we have
\begin{equation}\label{moret}
\begin{aligned}
&h_q(q,p)=\sum_{n=0}^\infty 2(n+1)a_{2(n+1)}(p)q^{2n+1},\\[1ex]
& h_p(q,p)=\sum_{n=0}^\infty a_{2n}'(p)q^{2n},\\[1ex]
& h_{pp}(q,p)=\sum_{n=0}^\infty a_{2n}''(p)q^{2n},\\[1ex]
& h_{pq}(q,p)=\sum_{n=0}^\infty  2(n+1)a_{2(n+1)}'(p)q^{2n+1},\\[1ex]
&   h_{qq}(q,p)=\sum_{n=0}^\infty 2(n+1)(2n+1)a_{2(n+1)}(p)q^{2n}.
\end{aligned}
\end{equation}
We now use  Cauchy's product formula for series to infer from \eqref{moret} that   for all~${|q|<L^{-1}}$ and $p\in[\ov p,0]$  we have
\begin{equation*}
\begin{aligned}
\big(h_q^2 h_{pp}\big)(q,p)&=\sum_{n=1}^\infty\bigg[\sum_{k=1}^n a_{2(n-k)}''(p)\sum_{\ell=1}^{k}4\ell(k-\ell+1)a_{2\ell}(p)a_{2(k-\ell+1)}(p)\bigg]q^{2n } ,\\[1ex]
\big(2h_ph_qh_{pq}\big)(q,p)&=8\sum_{n=1}^\infty  \bigg[\sum_{k=0}^{n-1} a_{2(n-k-1)}'(p)\sum_{\ell=1}^{k+1}\ell(k-\ell+2)a_{2\ell}(p)a_{2(k-\ell+2)}'(p)\bigg]q^{2n} ,\\[1ex]
\big(h_p^2h_{qq}\big)(q,p)&=2\sum_{n=0}^\infty  \bigg[\sum_{k=0}^n(k+1)(2k+1)a_{2k+2}(p) \sum_{\ell=0}^{n-k}a_{2\ell}'(p)a_{2(n-k-\ell)}'(p)\bigg]q^{2n} ,\\[1ex]
\gamma(p) h_p^3(q,p)&=\sum_{n=0}^\infty \gamma(p)\bigg[\sum_{k=0}^na_{2k}'(p)\sum_{\ell=0}^{n-k}a_{2\ell}'(p)a_{2(n-k-\ell)}'(p)\bigg]q^{2n}.
\end{aligned}
\end{equation*}
Plugging  these expressions and \eqref{moret}  for $h_{pp}$ into the quasilinear  equation  \eqref{eq:hod}$_1$ we obtain, after identifying for each fixed $p\in[\ov p,0]$ the coefficients of each power of $q$ the following identities
\begin{subequations}\label{FFF}
\begin{equation}\label{FFFa}
\begin{aligned}
a_2&=\frac{\gamma a_0'}{2}-\frac{a_0''}{2a_0'^2},
\end{aligned}
\end{equation}
and 
\begin{equation}\label{FFFb}
\begin{aligned}
a_{2(n+1)}&=\frac{1}{2(n+1)(2n+1)a_0'^2}\bigg\{-a_{2n}''-\sum_{k=1}^n a_{2(n-k)}'' \sum_{\ell=1}^{k}4\ell(k-\ell+1)a_{2\ell} a_{2(k-\ell+1)} \\[1ex]
&\hspace{4.2cm}+8\sum_{k=0}^{n-1} a_{2(n-k-1)}' \sum_{\ell=1}^{k+1}\ell(k-\ell+2)a_{2\ell} a_{2(k-\ell+2)}' \\[1ex]
&\hspace{4.2cm}-2\sum_{k=0}^{n-1} (k+1)(2k+1)a_{2k+2} \sum_{\ell=0}^{n-k}a_{2\ell}' a_{2(n-k-\ell)}' \\[1ex]
&\hspace{4.2cm}+\gamma\sum_{k=0}^na_{2k}'\sum_{\ell=0}^{n-k}a_{2\ell}'a_{2(n-k-\ell)}'\bigg\},\qquad n\geq1.
\end{aligned}
\end{equation}
\end{subequations}
in $[\ov p,0]$.
We note that, in view of \eqref{Eq3}, we have 
$$a_0'(p)=h_p(0,p)=-\frac{1}{u\circ \cH^{-1}(0,p)}>0 \qquad\text{for all $p\in[\ov p, 0].$}$$  
Consequently, the formulas \eqref{FFF} are well-defined.
Moreover, it is important to point out that the right side of \eqref{FFFb} is expressed only in terms of derivatives of $\gamma$  and  $a_{2k} $ with~${k\leq n}$.
We may thus conclude that all the coefficients $a_{2n}$ with $n\geq 1$ in the Taylor series of $h$ are determined by $a_0$ and $\gamma$.  
Thus, knowing $\gamma$ and $a_0$, we may recover the function~$h$  in the rectangle $(-L^{-1},L^{-1})\times[\ov p,0].$
By analytic continuation (note that the formula~\eqref{Taylor} is valid for all $q_0\in\R$) this determines $h$ in the strip $\R\times[\ov p,0],$ hence the wave surface and the fluid velocity in the surface layer 
 $\{(x,y)\in\ov{\0_\eta}\,:\, 0\leq \psi(x,y)\leq -\ov p\}$.

 In the final part of the proof we show how  $a_0$ can be determined from~${\mathfrak{u}(y)=u(0,y)}$ with~${y\in[\ov y,\eta(0)]}$.
 \begin{itemize}
 \item[(i)]  Recalling the definition of $\psi,$ we have
 \[\psi(0,y)=-\int_y^{\eta(0)}\mathfrak{u}(s){\rm d}s,\qquad y\in[\ov y,0]\]

 and 
 \[
\ov p=-\psi(0,\ov y)= \int_{\ov y}^{\eta(0)}\mathfrak{u}(s){\rm d}s.
 \]
 \item[(ii)]
 The function $-\psi(0,\cdot):[\ov y,\eta(0)]\to[\ov p,0]$ is in view of \eqref{Eq3} invertible.
Its inverse is exactly the function $a_0:[\ov p,0]\to [\ov y,\eta(0)]$.
 \end{itemize}
In view of (i)-(ii) we have identified  $a_0$ in terms of $\mathfrak{u}$, and this completes the proof. 
\end{proof}

\section{Some examples}\label{Sec:3}

In this last section we illustrate the recovery formula in the context of two famous explicit solutions for water waves of infinite depth.
Namely, we first  consider in Section~\ref{S:SS31}  Gerstner's wave (with and without stagnation points), which is a rotational periodic gravity wave, and in Section~\ref{S:SS32} Crapper's wave, which is a periodic irrotational capillary wave.
These examples illustrate both that the recovery formula provides better results if the line of symmetry is chosen to be a trough line.
This is in accordance to the key feature of nonlinear periodic water waves of having long flat troughs and short sharper crests, cf. e.g. \cite{Con11, SP88},
 which justifies a better convergence of the Taylor series, derived by our recovery formula, close to the wave trough.
 
\subsection{Recovery of Gerstner's wave}\label{S:SS31}

Gerstner's wave describes  periodic gravity waves that propagate  at the surface of a fluid of infinite depth. It is described in a Lagrangian framework
via the diffeomorphism $\Phi(t):=(X(t),Y(t)):\R\times (-\infty,b_0)\to\0_{\eta(t)}$ defined by
\begin{equation}\label{Lag}
 (X(t, a, b), Y(t, a, b)):=\left(a-\frac{e^{kb}}{k}\sin(k(a-ct)), b+\frac{e^{kb}}{k}\cos(k(a-ct))\right),\quad t\in\R,
\end{equation} 
where $b_0\leq0$, see~\cite{Ge09, He08}.
The constant $k>0$ is the wavenumber, $g$ is the  Earth's gravity, and   the wavespeed $c$ is given by
\[
c=\sqrt{\frac{g}{k}}.
\]
We now set $k=1$ and $g=1$, so that also $c=1$.
At time $t=0$ the wave surface is then the graph of the smooth curve~${(X(0, a, b_0), Y(0, a, b_0)):\R\to\R}$ with
\begin{equation*}
 (X(0, a, b_0), Y(0, a, b_0)):=\left(a-e^{b_0}\sin(a), b_0+e^{b_0}\cos(a)\right),
\end{equation*} 
which is a trochoid if  $b_0<0$ and a cycloid if  $b_0=0$  (and has cusps, thus  the regularity assumption \eqref{RA} is not satisfied in this case). 
One of the wave crests is located at $(x,y)=(0, b_0+e^{b_0})$.
At time~${t=0}$, the velocity of the fluid particles is given by
\[
(U(0,a,b),V(0,a,b))=\left( e^{b}\cos(a)),e^{b}\sin(a))\right).
\]
 We also note that the vertical halfline $\{(0,b)\,:\, b\leq b_0\}$ is mapped by $\Phi(0) $ onto the crest line $\{(0,b+e^{b})\,:\, b\leq b_0\},$ 
 which is a line of symmetry for the wave.
 
\begin{figure}[h!]
$$\includegraphics[scale=0.85, angle=0]{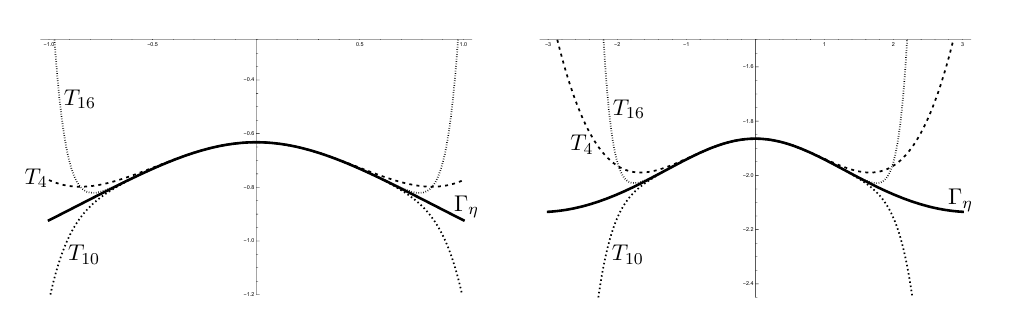}$$
\caption{Gerstner's wave $\Gamma_\eta$ close to the crest line~$\{x=0\}$  in the particular case~${k=1}$ and $g=1$ for the values~${b_0=-1}$ (left) and $b_0=-2$ (right) and the Taylor polynomials 
$T_{N}(q):=\sum_{n=0}^{N/2}a_{2n}(0)q^{2n},\, N\in\{4,\, 10,\, 16\}.$ 
}
\label{Figure1}
\end{figure}

 Let $\phi:(-\infty,b_0]\to (-\infty, b_0+e^{b_0}]$ be the diffeomorphism defined by $\phi(b)=b+e^b. $
 The horizontal velocity of the particle $\Phi(0,0,b)$ is given by $U(0,0,b)= e^{b},$
hence
\[
\mathfrak{u}(y)=e^{\phi^{-1}(y)}-1,\qquad y\leq b_0+e^{b_0},
\]
is the horizontal velocity of the fluid particles on the line of symmetry (in the moving frame).
Therefore, we have
\[
-\psi(0,y)= \int_y^{b_0+e^{b_0}}\big(e^{\phi^{-1}(s)}-1 \big)\,{\rm d}s,  \qquad y\leq b_0+e^{b_0}=\eta(0), 
\]
and ${a_0=(-\psi(0,\cdot))^{-1}}.$
Finally,  the vorticity  at $\Phi(0,b)=(0,\phi(b))$ is given by
\[
\omega(0,\phi(b)) = \frac{-2e^{2b}}{1-e^{2b}},\qquad b\leq b_0,
  \]
hence
\[
\omega(0,y) = \frac{-2e^{2\phi^{-1}(y)}}{1-e^{2\phi^{-1}(y)}},\qquad y\leq \eta(0),
  \]
  and 
\[
\gamma(p)=\omega(a_0(p))=\frac{-2e^{2\phi^{-1}(a_0(p))}}{1-e^{2\phi^{-1}(a_0(p))}},\qquad p\leq 0.
\]
Similar formulas are available for $\phi$, $\mathfrak{u}$, $a_0$, and $\gamma$ when considering the trough line $\{x=\pi\}$ as a line of symmetry.
In Figure~\ref{Figure1} we recover the wave profile in a neighborhood of the wave crest.

\vspace{-0.75cm}
\begin{figure}[h]
$$\includegraphics[scale=0.85, angle=0]{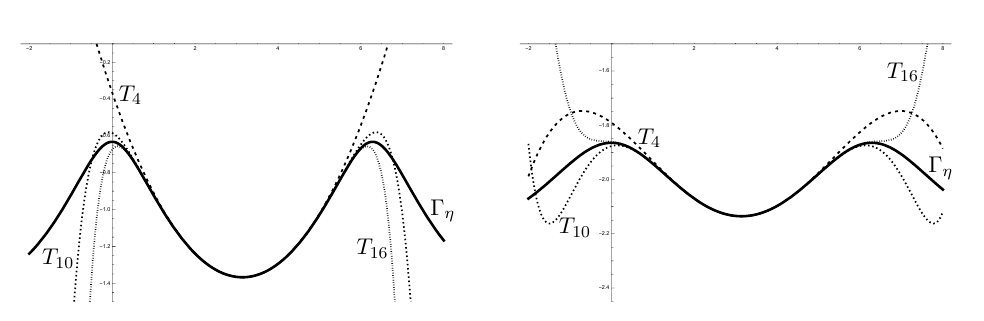}$$
\caption{Gerstner's wave $\Gamma_\eta$ close to the crest line~$\{x=\pi\}$  in the particular case~${k=1}$ and $g=1$ for the values~${b_0=-1}$ (left) and $b_0=-2$ (right) and the Taylor polynomials 
$T_{N}(q):=\sum_{n=0}^{N/2}a_{2n}(0)(q-\pi)^{2n},\, N\in\{4,\, 10,\, 16\}.$
}
\label{Figure2}
\end{figure}

In Figure \ref{Figure2} we have verified our recovery formula in a neighborhood of the wave trough $\{x=\pi\},$ which is also a line of symmetry for the flow. 

\vspace{-0.75cm}
\begin{figure}[h]
$$\includegraphics[scale=1, angle=0]{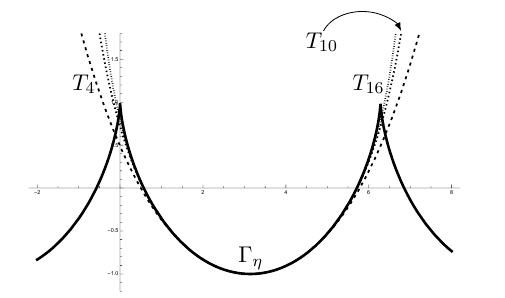}$$
\caption{Gerstner's wave $\Gamma_\eta$ close to the crest line~$\{x=\pi\}$  in the particular case~${k=1}$ and $g=1$ for the values~${b_0=0}$ and the Taylor polynomials 
$T_{N}(q):=\sum_{n=0}^{N/2}a_{2n}(0)(q-\pi)^{2n},\, N\in\{4,\, 10,\, 16\}.$ 
}
\label{Figure0}
\end{figure}

 Figure \ref{Figure0} illustrates that the recovery formula can be used also when stagnation points are present, but they do not lie on the line of symmetry.
 Indeed, if $b_0=0$, then Gerstner's wave has a stagnation point at the crest and therefore we choose again the wave trough $\{x=\pi\}$ as the line of symmetry of the flow.

The Taylor polynomials $T_4,\, T_{10},\, $ and $T_{16}$ appear to be a better approximations for the wave profile compared to the case when the line of symmetry is a crest line.  
The coefficients~$a_{2n}(0)$,~$n\leq 8$, in Figure~\ref{Figure1}-Figure~\ref{Figure0} are computed by using the recurrence formulas~\eqref{FFF} with the help of Mathematica. 

\subsection{Recovery of Crapper's wave}\label{S:SS32} 
Crapper's wave is a two-dimensional irrotational  periodic capillary wave (gravity effects are neglected) that propagates  at the surface of a fluid of infinite depth. 
The wave surface is the curve
\begin{equation*} 
\Big[a\mapsto\lambda\Big(a-\frac{2}{\pi}\frac{A\sin(2\pi a)}{1+2A\cos(2\pi a) +A^2},\frac{2}{\pi}-\frac{2}{\pi}\frac{1+A\cos(2\pi a)}{1+2A\cos(2\pi a) +A^2}\Big)\Big]:\R\to\R^2,
\end{equation*} 
cf. \cite{C57, CMa14,OS01}, where $\lambda$ is the wavelength.
The wavelength $\lambda$ is related to the parameter $A$, to the surface tension coefficient $\sigma$ (we recall that the density is set to be $\rho=1$), and the wavespeed $c$ via
\[
\lambda=2\pi\frac{1-A^2}{1+A^2}\frac{\sigma}{c^2}.
\]
Crapper's wave can be described as a graph if and only if $|A|<\sqrt{2}-1,$ see \cite{OS01}.
The flow corresponding to these wave is described by the equation
\begin{equation}\label{E:Flow}
z=\frac{w}{c}+\frac{2\lambda i}{\pi}\bigg(1-\frac{1}{1+Ae^{\tfrac{2\pi wi}{c\lambda}}}\bigg),
\end{equation}
\begin{figure}[h!]
$$\includegraphics[scale=0.85, angle=0]{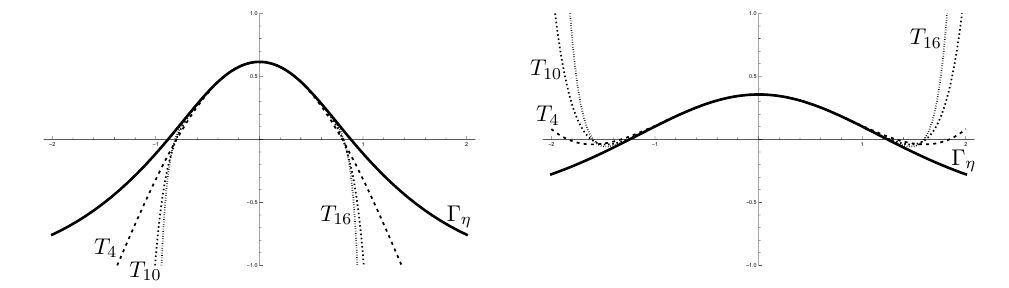}$$
\caption{Crapper's wave $\Gamma_\eta$ close to the crest line~$\{x=0\}$ for the values~${A=1/5}$ and $\lambda=(24\pi)/13$ (left) and~${A=1/10}$ and $\lambda=(198\pi)/101$ (right) together with the Taylor polynomials 
$T_{N}(q):=\sum_{n=0}^{N/2}a_{2n}(0)q^{2n} $ with~$N\in\{4,\, 10,\, 16\}.$ }
\label{Figure3}
\end{figure}
where $z=(x,y)\in\ov{\0_\eta}$ is an arbitrary point and $w=(\phi(z),\psi(z))$, with $\phi$ being the velocity potential, which satisfies $\nabla \phi=(u,v)$ and $\phi=0 $ on the crest line $\{x=0\}$, while $\psi$ is the stream function. Evaluating on the crest line $\{x=0\}$, we infer from \eqref{E:Flow} that $\psi(0,\cdot)$ satisfies the equation
\[
y=\frac{\psi(0,y)}{c}+\frac{2\lambda}{\pi}\frac{A}{A+e^{\tfrac{2\pi}{c\lambda}\psi(0,y)}}, \qquad y\leq\frac{2\lambda A}{\pi(1+A)}. 
\] 
In particular, we find an explicit expression for $a_0$, namely that
\[
a_0(p)=-\frac{p}{c}+\frac{2\lambda}{\pi}\frac{A}{A+e^{-\tfrac{2\pi}{c\lambda}p}}, \qquad p\leq0. 
\] 
A similar formula is available for $a_0$ when considering the trough line $\{x=\lambda/2\}$ as a line of symmetry.
We illustrate the recovery method for this wave in the particular case when
\[
\qquad c=1\qquad\text{and}\qquad \sigma=1.
\]

\vspace{-0.25cm}
\begin{figure}[h!]
$$\includegraphics[scale=0.85, angle=0]{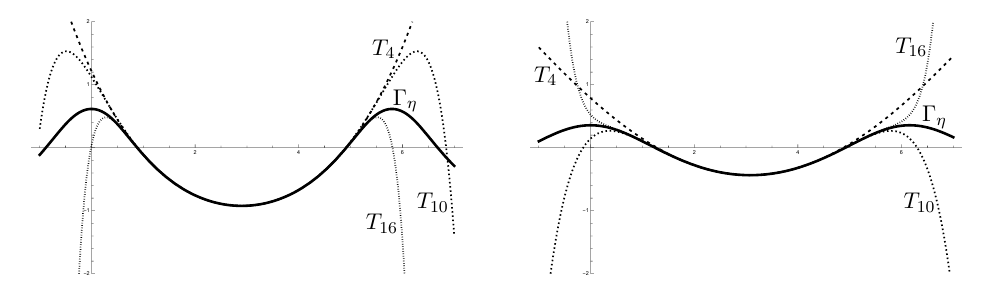}$$
\caption{Crapper's wave $\Gamma_\eta$ close to the crest line~$\{x=\lambda/2\}$ for the values~${A=1/5}$ and $\lambda=(24\pi)/13$ (left) and~${A=1/10}$ and $\lambda=(198\pi)/101$ (right) together with the Taylor polynomials 
$T_{N}(q):=\sum_{n=0}^{N/2}a_{2n}(0)\Big(q-\frac{\lambda}{2}\Big)^{2n}$  with $ N\in\{4,\, 10,\, 16\}.$ }
\label{Figure4}
\end{figure}
In Figure \ref{Figure3} we recover Crapper's wave in a neighborhood of the  crest line~${\{x=0\}}$ and in Figure \ref{Figure4} we  consider the trough line~${\{x=\lambda/2\}}$ as a line of symmetry.
Similarly as in the case of Gerstner's wave, the Taylor polynomials $T_4,\, T_{10},\, $ and $T_{16}$   approximate better the wave profile when the line of symmetry is a trough  line.   
Also in Figure~\ref{Figure3} and Figure~\ref{Figure4} we have computed the coefficients $a_{2n}(0)$, $n\leq 8$, by using the recurrence formulas~\eqref{FFF} with the help of Mathematica. 

\bibliographystyle{siam}
\bibliography{Literature}
\end{document}